\documentclass[doublecol,11pt]{article} % for 2 columns style without line
\usepackage{amsfonts}
\usepackage{graphicx}% Include figure files
\usepackage{dcolumn}% Align table columns on decimal point
\usepackage{bm}% bold math
\usepackage{amsmath}

\usepackage{graphicx}% Include figure files
\usepackage{dcolumn}% Align table columns on decimal point
\usepackage{bm}% bold math
\newcommand{\eps}{ { \varepsilon } }
\renewcommand{\phi}{{\varphi}}

\newcommand{\bk}{\vett k}
\newcommand{\bh}{\vett h}

\newcommand{\vett}[1]{\mathbf{#1}} 
\DeclareMathOperator{\im}{\mathrm{Im}}
\makeatletter
    \renewcommand*{\@fnsymbol}[1]{\ensuremath{\ifcase#1\or \textrm{a}\or
        \textrm{b}\or \textrm{c}\or \mathsection\or \mathparagraph\or \|\or
        **\or \dagger\dagger  \or \ddagger\ddagger \else\@ctrerr\fi}}
\makeatother

\title{Classical  infrared  spectra of ionic crystals
and their  relevance for statistical mechanics}
\author{Andrea Carati\thanks{Dep. Mathematics, Universit\`a degli
    Studi di Milano, Via Saldini 50, 20133 Milano -- Italy.}  
  \and   Luigi Galgani\footnotemark[1]%\inst{1} 
  \and Alberto Maiocchi\footnotemark[1]%\inst{1} 
  \and Fabrizio Gangemi\thanks{DMMT, Universit\`a di Brescia, Viale
    Europa 11, 25123 Brescia -- Italy.} 
  \and Roberto Gangemi\footnotemark[2]%\inst{2} 
}
\date{\today}% It is always \today, today,
\begin{document} 

\maketitle
\begin{abstract}
It was recently shown that the experimental infrared spectra of ionic
crystals at room temperature are very well reproduced by classical
realistic models, and here new results are reported on the temperature
dependence of the spectra, for the LiF crystal.  The principal aim of
the present work is however to highlight the deep analogy existing
between the problem of spectra in ionic crystal models on the one
hand, and that of energy equipartition in the Fermi--Pasta--Ulam
model, on the other.  Indeed at low temperatures the computations of
the spectra show that the dynamics of the considered system is not
completely chaotic, so that the use of the Boltzmann--Gibbs statistics
is put in question, as in the Fermi--Pasta--Ulam case.  Here, however,
at variance with the equipartition problem, a first positive
indication is given on the modifications that should be introduced in
a classical statistical treatment: the new results at low temperatures
show that it is indeed the Clausius identification of temperature that
has to be modified.  In fact, at very low temperatures the theoretical
spectra fail to reproduce the experimental ones, if the temperature is
taken as proportional to mean kinetic energy, but agreement is
recovered through the only expedient of introducing a suitable
temperature rescaling.  Analogous results are also found in connection
with thermal expansion.

\vskip 1.em
\noindent \textbf{Keywords}: Infrared spectra, equipartition principle,
ordered and chaotic motions, ionic crystal model, FPU model

\end{abstract}

\section{Introduction}
The present paper reports new results concerning classical theoretical
estimates of infrared spectra of ionic crystals.  In two previous
papers (see \cite{lif1} and \cite{quarzo}) the estimates were given
for spectra at room temperature, and here the temperature dependence
of the estimates is investigated, particularly at low temperatures,
for the Lithium Fluoride (LiF) crystal.

\begin{figure}[t]
  \begin{center}
    \includegraphics[width = 0.75\textwidth]{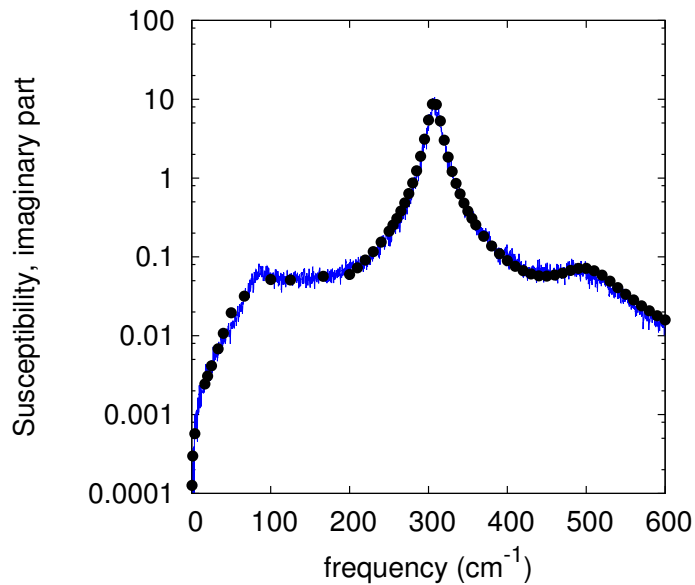}
  \end{center}
  \caption{Imaginary part of susceptibility vs frequency,  at room
    temperature. Comparison between calculations (solid line) and
    experimental data taken from \cite{palik} (points). Here, as in
    all following figures but Fig.~\ref{fig:r-lowt}, the computations are
    performed at a kinetic energy proportional to temperature according
  to the Clausius identification (\ref{clau}). }
 \label{fig:susc}
\end{figure}

\begin{figure}[t]
 \begin{center}
    \includegraphics[width =
      0.75\textwidth]{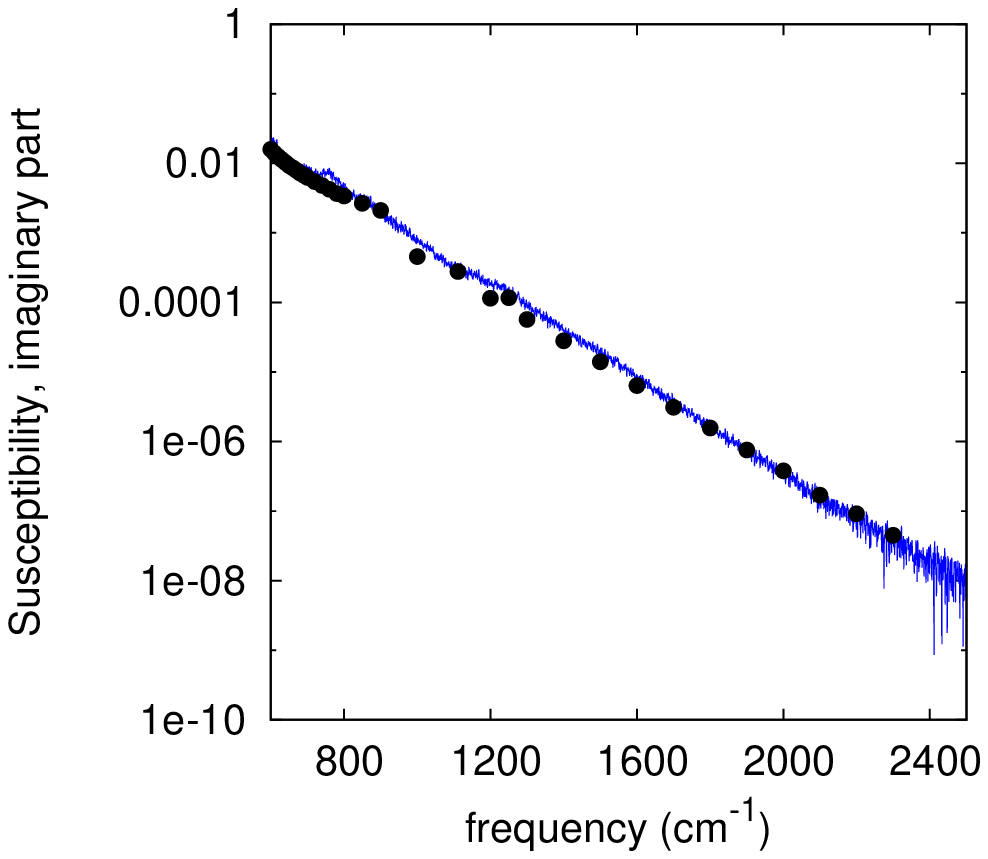}
  \end{center}
  \caption{High frequency behavior of the imaginary part of
    susceptibility vs frequency (at room temperature) 
compared with experimental data. }
 \label{fig:susc-hf}
\end{figure}

Thus stated, the problem seems to be one of interest for solid state
physics, which indeed is the case.  However, in this paper it is
pointed out that we are actually meeting here with quite general
problems of statistical mechanics.  First of all, we are meeting with
the problem of the relations between quantum statistical mechanics and
its classical counterpart, if not between quantum and classical
physics altogether. Because the theoretical spectra discussed here
reproduce well the experimental data, as shown by Fig.~\ref{fig:susc}
and \ref{fig:susc-hf} (for the LIF crystal at room temperature), while
they are computed in purely classical terms involving solutions of
Newton equations for the ions' motions, with no reference at all to
energy levels and corresponding jumps.

Furthermore, we are meeting with a problem concerning the dynamical
foundations of classical statistical mechanics. Indeed it will be seen
that, as temperature is diminished, the dynamics of the considered
model becomes less and less chaotic, so that the use of the
Boltzmann--Gibbs statistical mechanics becomes less and less
justified,  as in the FPU case.  On the other hand, a first positive
indication is also provided here, because it is shown that what should
be modified in a statistical treatment, is the Clausius identification
of temperature $T$ in terms of mean kinetic energy $\langle K\rangle$,
namely, the relation
\begin{equation}\label{clau}
\langle K\rangle =\frac 32 N k_BT \ ,
\end{equation}
where $N$ is the number of particles and $k_B$ the Boltzmann constant.
Indeed it will be seen that, using the Clausius identification, at low
temperatures the theoretical spectra fail qualitatively in connection
with a certain feature of the spectrum.  However, agreement is
recovered if, at each of the two low temperatures considered (85 K and
7.5 K), the initial data are taken at a suitable value of the mean
kinetic energy, which, through the Clausius identification, would
correspond to a much larger ``effective temperature" (180 and 125
K respectively). Analogous results are also obtained in connection
with the thermal expansion.

The plan of the paper is as follows.  In section \ref{sec2} we
illustrate some general features concerning realistic ionic crystal
models, in particular how they can be considered as evolutions of the
FPU model, and how ionic spectra can be dealt with in a classical
statistical mechanical frame.  In section \ref{sec3} we give details
on the concrete model used for the LiF crystal actually
investigated. In section \ref{sec4} we report the results on the
spectra. First, results at room temperature which extend previous
ones, and then their temperature dependence. Some comments are finally
given in section \ref{sec5}.  In an appendix some details are given
concerning the determination of the parameters of the model, making
use of the dispersion relations.

\section{Realistic ionic crystal models as evolutions of the FPU model. 
Classical  infrared  ionic spectra}\label{sec2}
It is well known that the FPU work put in question the applicability
of the Boltzmann--Gibbs statistics in connection with the
equipartition principle, for dynamical systems that don't present
sufficiently chaotic motions. Now, the failure of the equipartition
principle was observed in the FPU model, which is a simple, idealized
model of a one--dimensional crystal. So it is not clear whether the
FPU critique applies to concrete physical systems. On the other hand,
a simple natural generalization of that model exists which emulates
real systems, actually, ionic crystals. Moreover, ionic crystals are
endowed with dielectric properties, so that they present optical
spectra, which is a further physical phenomenon, perhaps the most
characteristic one of quantum physics, that they allow to investigate.
 
%In the present section we recall some general relevant features of
%ionic crystal models. First, how does it occur that optical spectra in
%the infrared can be described, and computed, in purely classical
%terms, involving ions' motions only, with no reference at all to
%energy levels and related jumps. Then, how the structure of the
%spectra reflects the chaoticity level of the considered dynamical
%system, so that the analogy with the FPU problem is made manifest.
%Taking into account that the spectra computed at room temperature
%reproduce in an impressive way the experimental data, one thus sees
%that ionic crystal models are realistic models of FPU--type, on which
%the applicability of the Boltzmann--Gibbs statistics can be tested,
%even in connection with a characteristic quantum phenomenon such as
%that of spectral lines.  We now briefly illustrate these facts.
 
Ionic crystals are known to be the simplest type of crystals (see
\cite{seitz}\cite{bh}).  For such crystals, as for the covalent
ones, the degrees of freedom of the electrons can be eliminated,
inasmuch as the electrons don't move around the crystal, but remain
trapped, each about its own ion.  Thus, at least for what concerns
certain physical phenomena, the crystal can be reduced to a system of
ions.  Indeed, it was shown by Born (and proved later with quantum
mechanical methods -- see for example \cite{eyring}) that the electrons
can just be assumed to produce on the ions two effects: first, to
endow each ion with a suitable ''effective charge'' entering the
mutual Coulomb forces (thus emulating the ions' polarizability), and,
second, to produce a further mutual repulsive ``effective short--range
force" among pairs of ions.  Born himself suggested for the potential
of the short--range force, the form $V^{(s)}(r)=a/r^6$, which involves
only one free parameter. Other more sophisticated effective potentials
were afterwards proposed, one of which was used in our model.  The
paradigmatic ionic crystal is that of LiF, which plays for crystals a
role analogous to that of Hydrogen atom for atomic gases.

At this point, having eliminated the electrons, the ionic crystal
model appears as a natural generalization of the FPU model, namely, as
a system of point particles oscillating about the sites of a regular
lattice, interacting through mutual two--body forces, which can be
studied in terms of normal modes referred to the minimum of the
potential energy. This was actually done in the year 1912 by Born and
von K\'arm\'an \cite{bvk} who, considering a model with two
alternating masses, could highlight the distinction between acoustic
and optical modes. Such a purely mechanical model, in its
one--dimensional version, became a prototype for studies in
perturbation theory related to the FPU model (see\cite{gg}), and is
presently investigated in the thermodynamic limit within the modern
statistical approach (see \cite{andrea1}), along the lines of
\cite{alberto} and \cite{deroeck}.  However, any realistic model of a ionic crystal has
to take into account also its dielectric properties, which are
described in terms of ionic polarization. This is given by
\begin{equation}\label{pola}
\vett P= \frac 1{V} \sum_{\bh,s} e_s \vett q_{s,\bh}\ ,
\end{equation}
where $\vett q_{s,\bh}$ is the displacement of ion $s$ in cell $\bh$
from its equilibrium position, $e_s$ its electric charge, $V$ the
volume.  In this way infrared spectra enter the game.

Optical spectra are thermodynamic properties of a system, just as heat
capacity and compressibility, which in statistical mechanical terms
are defined as responses to changes of an external parameter, here the
electric field. Optical spectra actually show up as \emph{spectral
  curves}, namely, functions which give the refractive index
$n(\omega)$ versus frequency, or the reflectivity $R(\omega)$ and so
on. All these are amenable to the curve $\chi(\omega)$ of electric
susceptibility, or more precisely of the corresponding tensor
$\chi_{ij}(\omega)$ in the crystal case.

The statistical mechanical expression for dielectric susceptibility
took a rather long time to be established, actually up to the end of
the years fifties, when linear response theory became available.  Some
papers by Gordon (see \cite{gordon}) reconstruct the rather painful
path that had to be followed. Starting from Schr\"odinger perturbation
theory for estimating probabilities of energy jumps, and then passing
to the Heisenberg picture, it was finally realized that the
statistical mechanical susceptibility formula can be expressed in
terms of essentially the time--autocorrelation function of polarization,
%$\vett P$, 
the latter being a microscopic quantity which, for the aims of
infrared spectra, involves the positions of the ions only.  Such a
formula is just a standard Green--Kubo one, which can be deduced also
in a purely classical frame (see ~\cite{andrea2}), and more precisely
has the form
\begin{equation}\label{kubo0}
\chi_{ij}^{ion}(\omega) = \frac V{\sigma^2_p}\int_0^{+\infty}
e^{-i\omega t} \langle P_i(t) \dot P_j(0)\rangle d t \ .
\end{equation} 
Here $\langle \ldots \rangle$ denotes ensemble average, $P_j$ the
$j$--th component of the ionic polarization $\vett P$ given by (\ref{pola}),
and $\sigma^2_p$ is a constant that reduces to $1/\beta$, if the
average is performed using the Gibbs ensemble. In this case the
formula takes the form
\begin{equation}\label{kubo}
\chi_{ij}^{ion}(\omega) = \frac{V}{k_BT}\int_0^{+\infty} e^{-i\omega
  t} \langle P_i(t) \dot P_j(0)\rangle d t \ ,
\end{equation} 
Anyhow, the ionic electric susceptibility turns out to be
expressed in terms of essentially the time--autocorrelation function 
of the ionic
electric polarization, a fact whose relevance will be pointed out
below.

The contribution of electrons to the total susceptibility tensor
$\chi_{ij}$ in the infrared amounts to a constant, which can be
conveniently written as $(\epsilon_\infty-1)/4\pi$, so that the
electric permittivity tensor, which is defined by $\epsilon_{ij}=
\delta_{ij}+4\pi \chi_{ij}$, takes the form
\begin{equation}\label{perme2}
\eps_{ij}(\omega)= \eps_\infty\delta_{ij}+4\pi
\chi_{ij^{}}^{ion}(\omega) .
\end{equation}

In the case of LiF, which will be considered in this paper, the
susceptibility tensor, and so also the permittivity tensor, are known
to be multiples of the identity, thus reducing to scalar
quantities. This property shows up also in the numerical computations,
as already observed in Ref.~\cite{lif1}. Thus we take for $\chi$ one
third the trace of the tensor $\chi_{ij}$.

Having available the statistical formula for electric permittivity,
one can eventually recover macroscopic optical quantities such as
refractive index $n$ and reflectivity $R$, which are given (see
\cite{born}) by
\begin{equation}
%\label{refl}
\nonumber n=\sqrt{\eps}\ ,\quad\quad
R=\left|\frac{\eps-1}{\eps+1}\right|\ .
\end{equation}

The statistical mechanical formula (\ref{kubo}) for electric
susceptibility is the clue for the possibility itself of a purely
classical approach to spectral lines.  The point is that in such a
formula any reference to energy levels and jump probabilities
completely disappeared, and the microscopic relevant quantity is the
time derivative of essentially the autocorrelation of the electric polarization,
which involves particles positions and momenta only. Now, the same
features occur also in the corresponding quantum formula, in which
however positions and momenta are Heisenberg noncommuting
operators. Thus, within a quantum approach the classical formula
(\ref{kubo}) is just understood as the corresponding \emph{classical
  approximation}, the advantage of which with respect to the quantum
formula is only that it can be concretely estimated through numerical
computations (i.e., through molecular dynamics (MD) simulations), the
same technique of the old FPU paper. But the purely classical formula
should just give a first approximation, to which suitable
\emph{quantum corrections} ought to be added, especially when working
in a fully quantum regime.
 
Indeed, some discussions are available in the literature (see
\cite{corrections}\cite{ciccotti}), concerning the choice of the best
quantum correction.  However, Figs.~\ref{fig:susc} and
\ref{fig:susc-hf} show that a simple naive implementation of the
purely classical formula for a model of LiF crystal, reproduces quite
well the experimental data at room temperature, i.e., well below the
Debye temperature (which is estimated to be about 730 K for the LiF
crystal), and thus in a deep quantum regime, with no need of any
quantum corrections at all.

We now come to the relations between infrared spectra and chaoticity.
The point is that in ergodic theory chaoticity can be defined as
corresponding to mixing, which is a property entailing that
 the time--correlation functions of all pairs
of dynamical variables vanish for sufficiently long times.
Now, 
the infrared spectrum is proportional to the Fourier transform of
essentially the
time--autocorrelation of electric polarization, and so the form of the
infrared spectrum is reflected in the chaoticity properties of the
considered dynamical system.  In particular, in the low--energy limit,
characterized  by non chaotic motions (actually almost periodic ones) 
a line spectrum (one in
which the imaginary part of dielectric susceptibility is a sum of
delta functions)  is expected to occur.  So 
the relevant  correlation functions are expected to
have relaxation times which  increase, even possibly diverging, as
temperature is decreased. To this end one can  compute the time
dependence of the time-- correlation function 
$\langle P_x(t) \dot P_x(0)\rangle$,  which can be conveniently normalized 
through division of $ P_x(t)$ and $ \dot P_x(0)$ by the corresponding 
standard  deviations.
\begin{figure}  
 \begin{center}
    \includegraphics[width = 1.\textwidth]{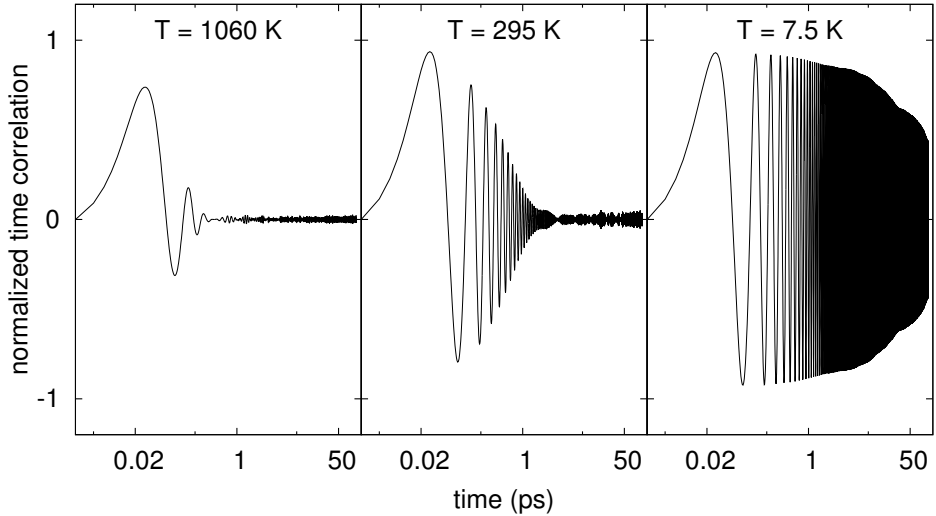}
  \end{center}
  \caption{The normalized correlation function $\langle P_x(t) \dot
    P_x(0)\rangle$ vs time at three
    simulation temperatures: 1060, 295 and 7.5 K, from left to
    right. Here, $P_x$ is the $x$ component of electric polarization.}
\label{fig:corr}
\end{figure}

In Fig.~\ref{fig:corr} we report the results obtained for the LiF model
studied in the present research, at three temperatures 
(defined through the Clausius
identification), namely, 1060, 295 and 7.5 K.
 One clearly sees that at the two higher temperatures 1060 K and  295 K 
 decorrelation  essentially occurs at $0.1$ ps and at $1$ ps
 respectively, whereas  at 7.5 K   decorrelation doesn't occur at all
 even at   100 ps, and much longer times seem to be required.  
This fact already indicates that some caution is necessary in applying 
the standard methods of classical statistical mechanics
at low temperatures.
A preliminary study of the decay time of the relevant correlation as a function
of temperature seems to hint at the  existence of  different laws for
high and low temperatures. This is a point that  we plan to better study in a
future work. The main goal of the present work is however to
concentrate on the form of the infrared spectra at low temperatures.

\section{The LiF crystal model}
\label{sec3}

The calculation of infrared spectra is reduced, as recalled above, to generating classical
orbits of the ions and performing suitable ensemble averages. We
accomplish this task by running MD simulations for a LiF crystal
model.

As in Ref.~\cite{lif1}, we represent the crystal as a system of $N$
point particles, with the masses of Li and F ions taken from
experiment. The interactions among the particles are described by a
two-body potential $V$, which includes both the Coulomb term and a
short range effective potential $V^{(s)}$,
\begin{equation}
\nonumber V(r_{ij})=\frac{e_ie_j}{r_{ij}}+V^{(s)}_{ij}(r_{ij})
\end{equation}
where indices $i,j$ denote the atomic species, Li or F, and
$e_i=\pm\ e_{eff}$.

For the short range effective potential, instead of the simple form
$V^{(s)}(r)=a/r^6$ proposed by Born and used in Ref.~\cite{lif1}, the
so--called Buckingham potential
\begin{equation}
\nonumber V^{(s)}_{ij}(r)=A_{ij}\exp(-B_{ij}r)-\frac{C_{ij}}{r^6}
\end{equation}
is adopted here.

A cubic simulation cell together with periodic boundary conditions is
used: long range electrostatic interactions are calculated by means of
a standard Ewald summation procedure, while for the short-range
effective forces a 5 \AA\ cut--off is applied.

The size of the simulation cell should be chosen at each temperature
in order to match the experimental density of the crystal (at ordinary
pressure).  At room temperature we take for the lattice step the
experimental value, i.e., 2.01 \AA\, as we did in Ref.~\cite{lif1}.
At different temperatures, instead, we adjust the size of the
simulation cell in such a way that the calculated primary peak shifts
in accordance with the experimental observations. As will be discussed
in detail later in connection with thermal expansion at low
temperatures, this procedure is pretty well compatible with the
experimental data for density.
\begin{table}[b]
 \begin{center}
   \begin{tabular}{cccc}
    \hline & $A$ (eV) & $B$ (\AA$^{-1}$) & $C$
    (eV$\cdot$\AA$^6$)\\ \hline Li-F & 3.30$\cdot 10^3$ & 5.00 &
    9.40\\ F-F\ & 15.8$\cdot 10^3$\ & 4.44\ & 41.6\\ Li-Li &
    3.18$\cdot 10^3$\ & 6.29\ & -2.77
  \end{tabular}
 \end{center}
 \caption{Optimized parameters of the short range potential with
   effective charge 0.7281 e.}
 \label{tab:table1}
\end{table}

Besides the lattice spacing, we have to fix the parameters of the
potential, namely, the effective charge $e_{eff}$ and the coefficients
$A_{ij},B_{ij},C_{ij}$ of the short range interaction.  To fix these
parameters, we first of all require that the stable equilibrium
position of the system corresponds to the LiF lattice, i.e., a FCC
structure with two ions per cell, with the two ions species
alternating along the three orthogonal directions. In addition, a fit
is made between the computed dispersion relations and the experimental
ones; this is explained in the Appendix. The parameters used (the same
for all temperatures) are reported in Table \ref{tab:table1}.

Finally, the choice of the parameter $\eps_\infty$ is done as follows.
Since the highest frequency we can reach, due to size of the
integration step, is about 8000 cm$^{-1}$, we require that
Eq. (\ref{perme2}) matches, at such a frequency, the experimental
value taken from Ref.~\cite{palik}.  In such a way, the value
$\eps_\infty=1.92$ is obtained. The choice of this parameter strongly
influences the decay of reflectivity on the side of high frequencies,
and in particular a value lower than 1.92 would give a better
agreement with the data.  We however chose the value 1.92 which is in
good agreement with the values reported in the literature for
$n_\infty$ of LiF (see Ref.~\cite{handbook}).

MD simulations are performed on a system of $N=4096$ ions. Numerical
integration of the equations of motion is performed at constant energy and
constant volume, using the Verlet
algorithm with an integration step of 2 fs. 

A key point concerns the energy that should be used in each
computation, namely, how should it be related to the experimental
temperature $T$.  We started our research using the procedure commonly
employed in MD simulations, which fixes the initial mean kinetic
energy $\langle K\rangle$ through the Clausius prescription
(\ref{clau}).  This can be achieved in several ways, and we just chose
the most direct one, sometimes called the {\it a posteriori} method.
Namely, in order to simulate the system at a temperature $T$ through
the corresponding kinetic energy $K$, the lattice is initially set at
its equilibrium configuration, and the ions are given random
velocities according to the Maxwell-Boltzmann distribution with a
certain temperature parameter $T_{in}$ (typically $T_{in}\simeq 2T$).
After a short transient time of 1000 integration steps, the lattice
temperature is calculated from the mean kinetic energy of the ions
and, if it differs from its target value $T$ by more than a threshold
amount, the process is repeated with a rescaled initial parameter
$T'_{in}$.  When the mean kinetic energy attains the desired value,
the actual simulation is started up, usually with a duration of 200
ps.

Ensemble averages in Eq. (\ref{kubo}) are replaced by a time average
over the simulation time, but in order to increase the statistical
significance of the results, multiple simulations (usually 10) are run
for each case.

\section{Results}
\label{sec4}

We preliminarily checked the method by working at room temperature.
The first two results concern $\im \chi$, the imaginary part of
susceptibility. The first one concerns its trend near the main peak,
which is reported in Fig.~\ref{fig:susc}. This figure can be compared
both with the results of our previous work Ref.~\cite{lif1}, and with
the old quantum estimates of Ref.~\cite{mara,wallis}. The improvement
with respect to our results of Ref.~\cite{lif1} may be attributed to
the larger number of parameters of the short range effective potential
and to the fact that the parameters are optimized by fitting the
dispersion relations. In particular the agreement with experiment
obtained at the two shoulders is indicative of a rather satisfactory
behavior of the non-linear part of the potential. An analogous
improvement occurs also for the real part of susceptibility, not
reported here.

The second result concerns the decay of $\im \chi$ for high
frequencies. As is well known, the experimental data (see
Ref.~\cite{kachare}) show that such a quantity becomes extremely small
at large frequencies, displaying an exponential decay over six orders
of magnitude. Thus $\im\chi$ becomes very small, and hard to estimate
numerically in a reliable way (see for example Ref. \cite{berens},
page 4873). This difficulty can be overcome, and in
Fig.~\ref{fig:susc-hf} the numerical results thus obtained are
compared with the experimental data, exhibiting an impressive
agreement over six orders of magnitude.

In the following lines, some technical details are discussed,
concerning two sources of error that had to be taken into account. The
first one is the loss of analyticity of the correlations, which has a
direct impact on the exponential decay (as explained in
Ref.~\cite{CaratiMaiocchi}), and is due to the truncation of the
integral (\ref{kubo}) to a finite time domain.
\begin{figure}[th]
  \begin{center}
    \includegraphics[height = 0.75\textwidth]{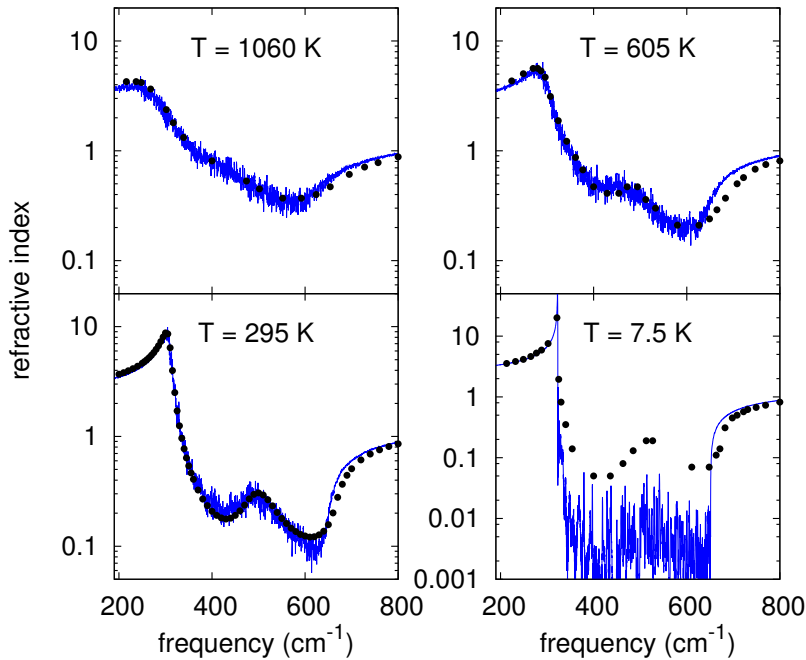}
  \end{center}
  \caption{Calculated refractive index spectrum (solid lines) compared
    with experimental data \cite{Li} (full circles), at several
    temperatures.}
 \label{fig:n-allT}
\end{figure}
The second source is the fact that the time average is performed over
a finite time domain, which implies that the correlations $\langle
P_i(t) P_i(0)\rangle$ are not positive definite in the Bochner sense,
as they have to be.  Analyticity is recovered by using a suitable
Gaussian filter. Instead, in order to minimize the distortions
introduced by the finiteness of the time average in the computation of
$\im\chi$, we perform the integration in (\ref{kubo}) in a symmetric
interval $[-t_f,t_f]$, by exploiting the parity property of the
correlations. Now, being known from theory that $\im\chi$ is positive
for positive frequencies, we consider as unreliable the spectrum in
the region where the computed values are negative, and this forces us
to restrict ourselves to the domain $\omega<2500$ cm$^{-1}$.

In order to investigate the temperature dependence of the spectra,
which is the main goal of the present work, we started considering the
refractive index.  In Fig.~\ref{fig:n-allT} the curves calculated at
several temperatures are compared with the experimental data taken
from Ref.~\cite{Li}.

\begin{figure}[t]  
 \begin{center}
    \includegraphics[width = 0.7\textwidth]{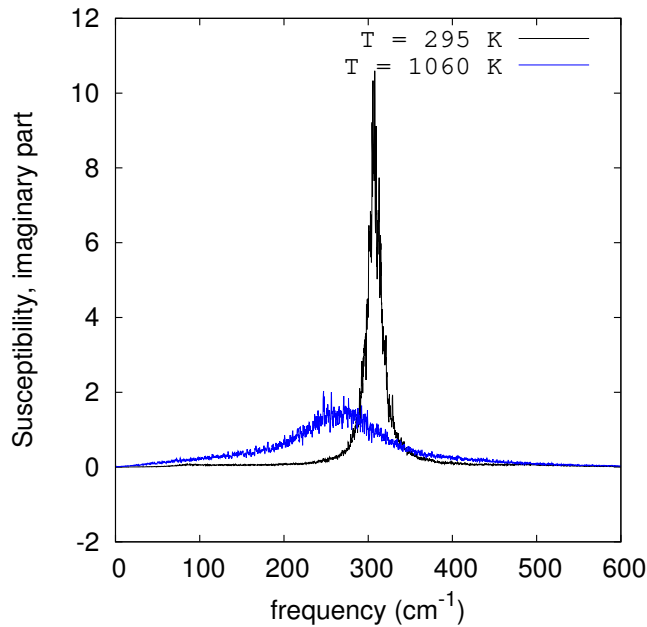}
  \end{center}
  \caption{Calculated susceptibility spectrum (imaginary part)
%as a function of   frequency 
at room temperature and at 1060 K (near the melting point).}
\label{im-chi-temp} 
\end{figure}
\begin{figure}[t] 
  \begin{center}
    \includegraphics[width = 1.0\textwidth]{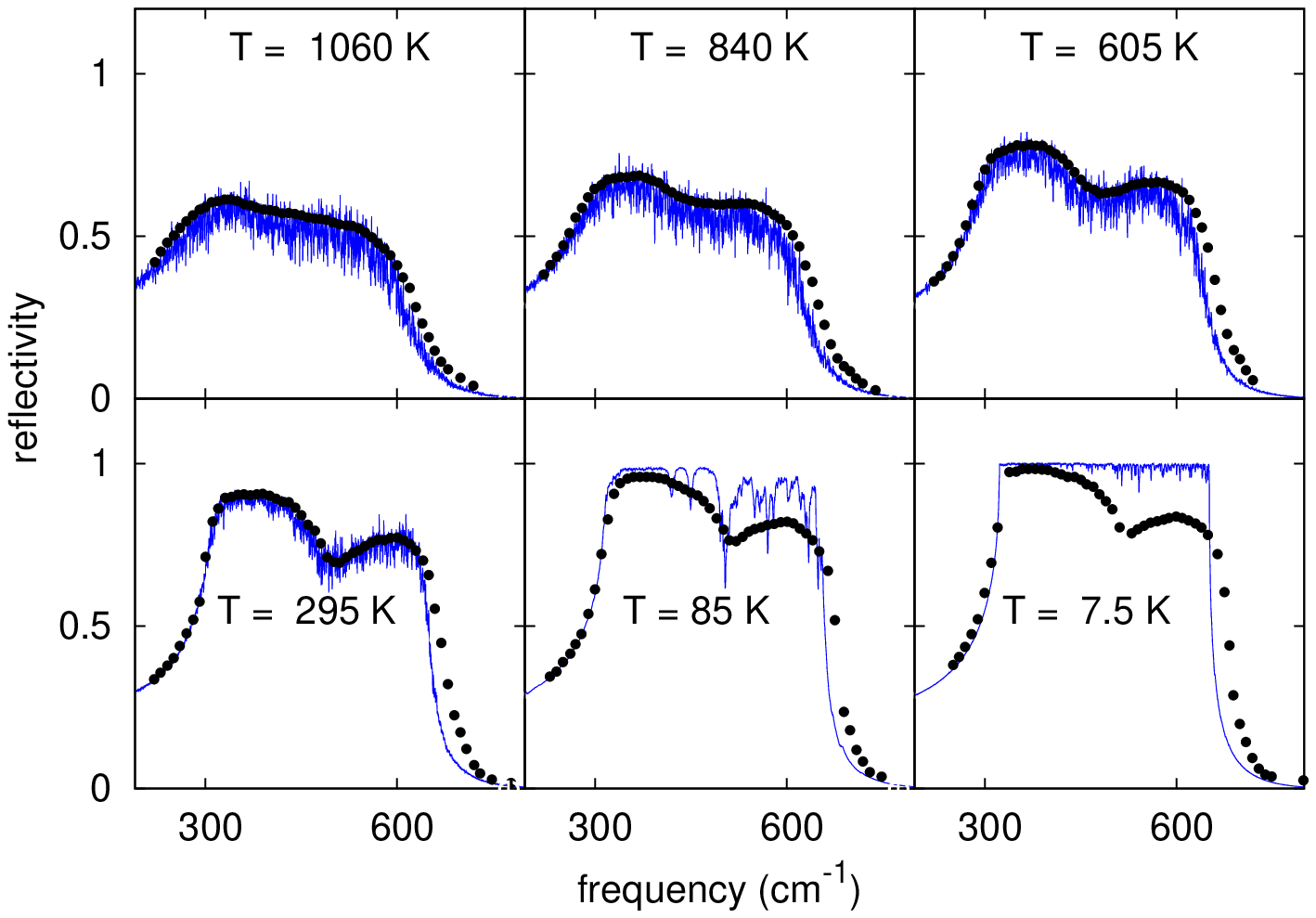}
  \end{center}
  \caption{Calculated reflectivity spectra  (solid line) compared with
     experimental data \cite{jasperse} (points) at several
     temperatures.}
  \label{r-alltnr}
 \end{figure}
Given the surprisingly good agreement of the classical spectrum with
the experimental one at room temperature (295 K), which is a
temperature lower than the Debye one (about 730 K for LiF), one may
expect that the agreement should even improve at higher temperatures.
In fact, at higher temperatures the calculated refractive index
matches the data nicely.  The spectra, sure enough, grow more noisy
and the peaks broaden, as can be better appreciated by looking
directly at the imaginary part of susceptibility
(Fig.~\ref{im-chi-temp}).  This is obviously due to the nonlinearity,
which increases as temperature is raised. The good agreement is a
proof of the ability of our potential to reproduce the non-linear
effects.  

Instead, it is apparent that at 7.5 K the calculated
spectrum substantially differs from the measured one in the region
between 300 cm$^{-1}$ and 600 cm$^{-1}$ (notice the  change of scale
for the ordinates, at 7.5 K), where a secondary peak is
present in the experimental data, but is  absent in the numerical
ones. Now, 7.5 K is just the temperature at which, as shown in 
Fig.~\ref{fig:corr} the relevant correlation did not yet decay, so that 
 one may wonder whether the discrepancy would
disappear  by performing  much longer computations exhibiting a full 
relaxation. It is clear however that  with longer computations the discrepancy,
rather than being eliminated, would even be enhanced. Indeed, dealing,
as we are doing here, with an integral truncated at a certain time
$\tau$, we are actually performing a  convolution between  the ``true'' 
Fourier transform and the test function $(\sin
\omega\tau)/\omega$.
Now, the effect of such a convolution with respect to a ``true'' peak
consists in smearing it out (and possibly adding some spurious peaks
very near the original one). Thus, a truncation at a larger time $\tau'>\tau$
would essentially produce an analogous  more pronounced peak   but not
a second  one at a sensible distance.
So, even if the correlation
function were computed up to a much longer time, possibly attaining
decorrelation, the computed spectrum would not reproduce the secondary peak exhibited by
the experimental data. The main contribution of the present paper consists, as
will be shown below, in pointing out that the remedy to this
discrepancy is obtained through a qualitatively new expedient, which involves
the identification of temperature in mechanical terms.

Concerning the discrepancy at 7.5 K,
the relevant point is that the frequency of the secondary peak
does not show up among the frequencies of the linearized system, so
that its presence should be an effect of the nonlinearity. So an
apparently contradictory situation occurs, because the
nonlinearity decreases with decreasing energy, whereas in the
experimental data the visibility of the secondary peak increases with
decreasing temperature.
\begin{figure}[t] 
  \begin{center}
    \includegraphics[width = 0.75\textwidth]{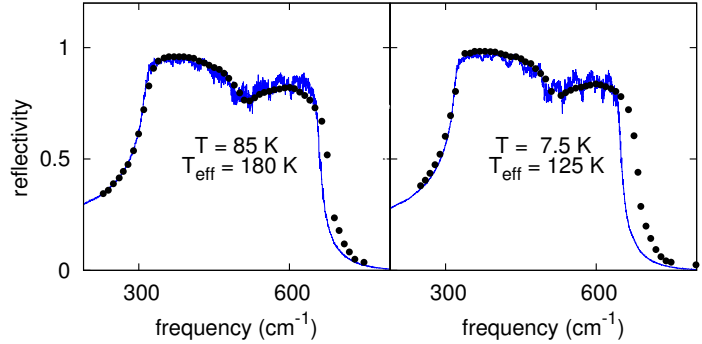}
  \end{center}
  \caption{Calculated reflectivity spectrum (solid lines) compared
    with experimental data \cite{jasperse} (points) at low
    temperatures, with temperature rescaling.  Here,
    $T$ refers to experimental data, while $T_{eff}$ is the rescaled
    value used in the computations.}
  \label{fig:r-lowt}
\end{figure}
%
%begin{figure}[t]
% \begin{center}
%   \includegraphics[height = 0.75\textwidth]{R-allT-2.eps}
% \end{center}
% \caption{Calculated reflectivity spectrum (solid line) compared with
%   experimental data \cite{jasperse} (points) at several
%   temperatures.}
% \label{fig:r-allT-risc}
%end{figure}

In order to better understand such a discrepancy, we decided to
consider also reflectivity,  for which a larger set of experimental
data is available (see for example Ref.~\cite{jasperse}), and which
was studied since a long time (see \cite{gottlieb}). Here one has to
take into account that the feature under discussion, namely, the occurring of
a secondary peak at
about 500 cm$^{-1}$ in the refractive index curves,  
manifests itself   in the reflectivity curves as a hollow. The
calculated reflectivity curves are reported, together with the experimental data,
in Fig.~\ref{r-alltnr}. One sees that
 in the experimental spectra the hollow starts becoming visible at 840 K, 
and its visibility increases as
temperature is lowered. Instead, the theoretical curves reproduce well
the experimental ones only from the highest temperatures down to room
temperature, because the hollow is poorly reproduced already
at 85 K, and completely disappears at 7.5 K, where the reflectivity
becomes essentially uniform, sticking at the value 1. 
\begin{table}[bh]
 \begin{center}
   \begin{tabular}{rrrr}
    \hline $T$ (K) &\ \ $a_0$(noresc) &\ \ $a_0$(resc)
    &\ \ $a_0$(exp)\\ \hline 7.5\ & 1.988 & 1.995 & 2.001\\ 85\ &
    1.995 & 2.001 & 2.002\\ 295\ & 2.010 & 2.010 & 2.010\\ 605\ &
    2.034 & 2.034 & 2.033\\ 840\ & 2.055 & 2.055 & 2.055\\ 1060\ &
    2.073 & 2.073 & 2.078\\
  \end{tabular}
 \end{center}
 \caption{Lattice steps used to reproduce data at several
   temperatures.  Second column: values used in our simulations with
   no rescaling; third column: values used in our simulations with
   rescaled temperatures; fourth column: values obtained by applying
   thermal expansion coefficient from Ref.~\cite{handbook} to the
   value at 295 K.}
 \label{tab:table2}
\end{table}

So, the  experimental results concerning the secondary peak (or the hollow) 
seem to  indicate that the
nonlinearity does not vanish as the temperature goes to zero, i.e.,
that  kinetic energy does not vanish at zero temperature.

Now, the nonvanishing of kinetic energy at zero temperature is nothing
but the well known Debye--Waller effect which, in quantum mechanical
terms, is just a manifestation of zero--point energy.  On the other
hand, in the present context in which the Boltzmann--Gibbs statistics
is put in question due to the non complete chaoticity of the motions,
the idea naturally presents itself that an analogue of zero--point
energy might be introduced in a classical model too. By the way, in
the literature it was already pointed out
 that quantum phenomena exist which can be emulated
by classical computations performed at suitable ``elevated''
temperatures. See Refs. \cite{ivanov},\cite{kumarmarx}, and
\cite{marxparrinello}, Fig. 2.

More concretely, having in mind the possibility of emulating
zero--point energy, and  without any
presumption of introducing a theory, in the present context we first
of all start abandoning the Clausius identification of temperature as
proportional to mean kinetic energy. Then, with a very simple,
pragmatic attitude, just by trial and error we look for two suitable
values of the mean kinetic energy which, inserted in the initial
data, may allow us to reproduce sufficiently well (if possible at all)
the experimental spectra corresponding to 85 and 7.5 K. Such values of
the kinetic energy can also be expressed in terms of ``effective
temperatures", just meant as the temperatures that correspond to given
kinetic energies through the Clausius prescription.  In such a way,
just by a lucky guess we
found the surprising result that the spectra at 85 and 7.5 K are
pretty well reproduced by classical computations at suitable effective
temperatures of 180 and 125 K respectively, as shown by
 Fig.~\ref{fig:r-lowt}. This is indeed the main,
unexpected result obtained in this paper.\footnote{By the way, it 
  turns out that, at the considered effective temperature of 125 K, the
  relevant correlation  actually decays to zero in about 10 ps. 
Thus  the problem of the truncation time which arises at an
effective temperature of 7.5 K  (see Fig.~\ref{fig:corr}) does not arise here.}
 
Apparently the agreement at the two low temperatures, both in
connection with the hollow
and in general for the whole spectrum, is of the same quality 
 as at the higher temperatures.  True, some disagreement is
observed at the right shoulder, but this occurs somehow at the same
degree at all temperatures. The reason of this minor fact is not yet
clear to us, and we hope to understand it in the future. Perhaps a
simple explanation might be related to the choice of the parameter
$\epsilon_\infty$ discussed in section \ref{sec3}.

A strictly related phenomenon involving temperature rescaling,
actually just the same rescaling as for spectra, was met in connection
with the lattice spacing. As explained in section \ref{sec3}, at all
temperatures different from 295 K the calculated spectra are obtained
by adjusting the lattice spacing in such a way that the position of
the main peak matches experiment.  The values of the lattice spacing
found in this way, with and without temperature rescaling, are
compared in Table \ref{tab:table2} with those obtained from the
experimental data available in the literature (see
Ref.~\cite{handbook}) for the thermal expansion coefficient of LiF.
The agreement above room temperature is impressive.  At 85 and 7.5 K,
instead, the agreement is not so good if temperatures are not
rescaled: indeed a qualitative failure is found here, because the
theoretical values continue to decrease almost linearly, while the
experimental ones exhibit a saturation effect. After temperature
rescaling, the value at 85 K perfectly agrees with the experimental
one, while at 7.5 K there is still a little deviation, but much less
than before.  This behavior corresponds to the one found in the
spectra, so that one is led to guess that the two effects stem from
the same origin.

\section{Conclusions}\label{sec5}
The results obtained are apparently interesting.  Indeed, already the
good agreement of the classical spectra with the experimental ones at
room temperature, obtained in a previous work
 without invoking any energy quantization,  
might have been unexpected. But the fact that even at
almost absolute zero the classical computations of the spectra
reproduce well the experimental data through the only expedient of
suitably rescaling temperature, all other purely classical ingredients
remaining unchanged, appears to be really surprising. Moreover, the
impression that here one is not meeting with some fortuitous
coincidence, is supported by the agreement found for thermal expansion
too. 

On the other hand it is well known that the Clausius identification of
temperature with kinetic energy is
in contrast with experiment, for example because zero--point energy
in solids exists.
This was indeed one of the main historical  reasons for abandoning
classical mechanics.  Here, however, it has been also shown that, for what
concerns infrared spectra, classical statistical mechanics is adequate
if a suitable rescaling of temperature is introduced.

So the problem arises whether the Clausius
identification of temperature 
 is a necessary implication in classical
statistical mechanics.  Some comments are given here.  
A first question in this connection is whether temperature should
be defined  as the mean value  of
some observable,   as is  suggested by the Clausius
identification of temperature with mean kinetic energy in the classical case. The question
then becomes of understanding which quantity does a thermometer measure if it doesn't
measure  kinetic energy. 
 On the other hand, we point out, at equilibrium temperature occurs, both in the
classical and the quantum cases, as a parameter 
which enters the Gibbs distribution, determining  the mean value of
energy. However, the temperature defined in such a way turns out to be
 proportional  to
mean kinetic energy  only in the classical case, while in the
quantum case  the relation between temperature and kinetic  energy 
is not universal, and  depends on the considered system.

So, not being the average of an observable, it is not clear how
temperature should be defined\footnote{Apart from introducing a
  microscopic model of a thermometer.} if the Gibbs ensemble is not
justified, as  occurs in  classical statistical mechanics in lack of
suitable ergodicity properties.  This is indeed
 the problem that was raised by the FPU work, which pointed
out that energy equipartition, the main consequence of classical
statistical mechanics, does not occur at low energies for the
simplest model of a crystal.  Apparently, the same occurs in
connection with the infrared spectra of ionic
crystals.
% and so  one meets with two cases of a same general 
%problem concerning   the foundations of classical statistical
%mechanics. In such a sense
%the  dynamical system studied here can be considered  to be
%essentially a realistic FPU--type model, with the only difference of
%being also    endowed
%with dielectric properties, which allow it to take  infrared spectra
%into account.

A provisional conclusion, in our opinion, might be the following.
The Clausius identification of temperature as mean kinetic energy
in classical statistical mechanics  seems not to be justified in
cases, such as that of FPU--type models at low energies, where suitable
ergodicity properties are not guaranteed. The general
 problem of how
a statistical mechanics should be formulated in such cases then
remains completely open, and actually is a formidable one. 
However the results illustrated here for  a realistic FPU--type model presenting
dielectric properties, with their surprising agreement with experiment,
seem to  indicate that
a general solution may exist.

\vskip .5 truecm

\textbf{Acknowledgments.}
The use of computing resources provided by CINECA is gratefully
acknowledged.
\begin{figure*}[th]
  \begin{center}
    \includegraphics[width = 1.04\textwidth]{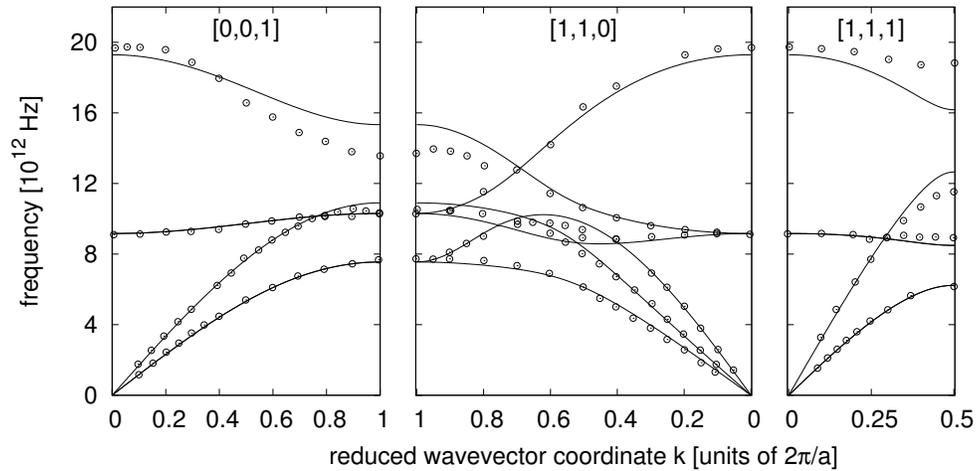}
  \end{center}
  \caption{Solid lines: calculated dispersion relations. Circles: data
    from Ref.~\cite{dolling} }
 \label{fig:disp}
\end{figure*}

\appendix

\section*{Appendix: fit of dispersion relations}
The parameters of the potential are determined by comparison of
calculated dispersion relations with experimental data taken from
Ref.~\cite{dolling}. At variance with MD simulations, we consider an
infinite crystal and linearize the model around the equilibrium
positions, obtaining an infinite number of linear equations, with
parameters which depend on the parameters of the potential. We then
find the normal modes solutions for such equations, i.e., solutions in
the form of travelling waves $\vett q_{s,\vett h}=\vett Q^{(l)}_s
e^{i(\bk\cdot \vett x_\vett h-\omega t)}$, thus determining the
dispersion relations $\omega=\omega^l(\bk)$ for the different
branches. Then we minimize the quantity $\sum_l\sum_i
|\omega^l(\bk_i)-\omega^l_i|^2$, $\omega^l_i$ being the experimental
values at $\bk=\bk_i$, by varying the parameters of the potential. The
best fit is shown in fig. \ref{fig:disp} for the three high symmetry
directions of the wave vector.  In general there are three optical and
three acoustical branches, two transverse and one longitudinal for
each group, with a possible degeneration in the transverse ones,
depending on the direction of $\bk$.  The agreement with data, which
is generally good for the acoustic branches and for the transverse
optical branches, is not so satisfactory for some parts of the
longitudinal optical branches, especially at high values of $|\bk|$.
This could be due to the absence of atomic polarizabilities in our
model: indeed, much better results were obtained in
Ref~\cite{dolling}, where atomic polarizabilities were also taken into
account in the model.

The optimized parameters we find are reported in
Table~\ref{tab:table1}.

\end{document}